    \documentclass[12pt,psf,epsf]{article}
\usepackage[centertags]{amsmath}
\usepackage{amssymb}
\usepackage{graphicx}
\usepackage{epsfig}
\usepackage{ulem}
\usepackage[english]{babel}
\usepackage{array}
\usepackage{amsthm}
\usepackage{latexsym}
\usepackage[mathcal]{euscript}
\pdfoutput=1
\usepackage{epsfig}

 \usepackage{jheppub}
 \usepackage{hyperref}
 \usepackage{amssymb}

\definecolor{darkraspberry}{rgb}{0.53, 0.15, 0.34}

\definecolor{darkblue}{rgb}{0., 0, 1}

\definecolor{dgreen}{rgb}{0.,0.6,0.}

\usepackage{subfigure}
\usepackage{psfrag}

\usepackage{epsfig}
\usepackage[latin1]{inputenc}
\usepackage{float}
\usepackage{graphicx}
\usepackage{cancel}
\usepackage{mathrsfs}
\usepackage{amssymb}
\usepackage{amsfonts}
\usepackage{amsmath}
\usepackage{slashed}
\usepackage{graphicx}
\usepackage{bm}
\usepackage{color}
\newcommand{\be}{\begin{equation}}
\newcommand{\ee}{\end{equation}}
\newcommand{\bea}{\begin{eqnarray}}
\newcommand{\eea}{\end{eqnarray}}
\newcommand{\bwt}{\begin{widetext}}
\newcommand{\ewt}{\end{widetext}}

\newcommand{\nn}{\nonumber}

\newcommand{\bi}{\begin{itemize}}
\newcommand{\ei}{\end{itemize}}

\usepackage{setspace}

\begin{document}

\title {On the nonclassicality in quantum JT gravity}

\author[a]{Dmitry S. Ageev,}
\author[a]{Irina Ya. Aref'eva}
\author[b]{and Anastasia V. Lysukhina}
\affiliation[a]{Steklov Mathematical Institute, Russian Academy of Sciences, Gubkin str. 8, 119991 Moscow,
Russia}
\affiliation[b]{Department of Particle Physics and Cosmology, Physics Faculty, M.V. Lomonosov Moscow State University,
Leninskie Gory 1-2, 119991 Moscow, Russia}
\emailAdd{ageev@mi-ras.ru, arefeva@mi-ras.ru,av.lysukhina@physics.msu.ru}

\abstract{In this note, we consider the question of classicality for the theory which is known to be the effective description of two-dimensional black holes - the Morse quantum mechanics. We calculate the Wigner function and the Fisher information characterizing classicality/quantumness of single-particle systems and briefly discuss further directions to study.
}

\maketitle

\newpage

\section{Introduction}
Quantum gravity and quantum information in their reciprocal relation are one of the central topics in high-energy theoretical physics at the moment \cite{Ryu:2006bv,VanRaamsdonk:2010pw,Swingle:2009bg}. The straightforward methods of higher-dimensional gravity quantization do not give a satisfactory result. At the same time, two-dimensional quantum gravity is the origin of many interesting results which could help in the understanding of its more complicated many-dimensional cousin. The canonical example of such a gravity theory is quantum Liouville theory, scalar field theory with exponentially growing potential \cite{DHoker:1982wmk}. Another interesting example is two-dimensional dilaton gravity which is especially favorable in the context of two-dimensional black holes \cite{Callan:1992rs} and which recently attracted a lot of attention in the context of information paradox \cite{Almheiri:2019psf,Maldacena:2016upp}. Also, it was found that the effective quantum dynamics of two-dimensional black holes reduces to some dual quantum system, namely quantum mechanics of single particle in exponentially growing potential \cite{Mertens:2017mtv,Mertens:2018fds}. The same quantum mechanical system describes the certain limit of boundary Liouville theory \cite{Ghoshal:1993tm,Dorn:2006ys,Dorn:2008sw}. In this way, we have the possibility to study the properties of such a mysterious object as a black hole describing the corresponding properties of Morse quantum mechanics\footnote{Typically by Morse quantum mechanics is called the case of confining potential with two exponentials. In the quantum gravity context the potential is usually unbounded and have continous spectrum. For a Fisher information and Wigner function considerations for a confining case see \cite{Habib:1990,Habib2,Chatterjee:2020,Hai-Woong:1982}}.

There are many quantities to characterize different unusual properties of the quantum world, which are interesting to study in the case of a single particle. In this note, we will focus on three quantities characterizing the interplay between classical and quantum behavior in the system. The first one is a Winger function (see \cite{rev1} for review and references) which in some sense characterizes quantum systems in both position and momentum spaces. It is well known, that the states like coherent ones. Wigner function is strictly positive, while for the complicated states where the quantum effects are of special importance it becomes negative for some values of position and momentum. In \cite{nonclas} it was proposed that the joint ``volume'' of the phase space where the Wigner function is negative is the measure of non-classicality corresponding to a certain system.

Another interesting measure which we consider in this note is the Fisher information which is the probabilistic quantity characterizing the measure of information contained in the random variable ${\cal X}$ about some unknown parameter $\theta$ in the distribution of ${\cal X}$. The Fisher information has quite a wide range of applications in quantum mechanics including the derivation of the Schrodinger equation from the first principles and measure of non-classically \cite{frieden,hall,MK}. Unfortunately, it is not clear how to apply the measures of nonclassicality related to Fisher information for potentials with continuous spectrum. By introducing the regularization in Morse potential we calculate Fisher information in different representations.
This note is organized as follows. In Sec.\ref{sec:intro} we set up the notation and introduce the relation between Morse quantum mechanics and gravity, in Sec.\ref{sec:W-WKB-Morse} we determine the Wigner function for it explicitly, and in Sec.\ref{sec:fisher} calculate Fisher information and non-classicality.

\section{Setup} \label{sec:intro}
\subsection{Two-dimensional gravity and Schrodinger equation}
Let us briefly list some models of two-dimensional quantum gravity that motivate the considerations of the quantum mechanical models in this note. The first one which attracted a lot of interest recently is a two-dimensional black hole in JT gravity given by the action
\bea 
S_{JT}=-\frac{1}{16\pi G}\left(\int d\tau dr \Phi\sqrt{g}(R+2)+2\int_{bdy} \Phi_{bdy}K \right),
\eea 
where $R$ is the scalar curvature of the metric, $K$ is extrinsic curvature along the cut-off surface at the boundary. The equations of motion for the metric
\be 
R=-2,
\ee 
fixes it from the very start and we choose it to be that of two dimensional black hole
\be 
d s^{2}=\left(r^{2}-1\right) d \tau^{2}+\frac{d r^{2}} {\left(r^{2}-1\right)}.
\ee 
We fix the wiggly cut-off along the boundary  which is parametrized as $(r(u), \tau(u))$ and choose the boundary condition to be  $\left.d s^{2}\right|_{\text {bdry }}=\Lambda^{2} d u^{2}$ and $\left.\Phi\right|_{\text {bdry }}=\Lambda \Phi_{r}$ for  large $\Lambda$. The extrinsic curvature around this cutoff surface is given by
\be 
K= \sqrt{r^{2}-1} 
 \cdot\left(\frac{\tau^{\prime}\left(r^{\prime \prime}+r\left(3 r^{\prime 2}+\left(r^{2}-1\right)^{2} \tau^{\prime 2}-r r^{\prime \prime}\right)\right)+\left(r^{2}-1\right) r^{\prime} \tau^{\prime \prime}}{\left(r^{\prime 2}+\left(r^{2}-1\right)^{2} \tau^{\prime 2}\right)^{3 / 2}}\right),
\ee 
and using the boundary condition one can express $r(u)$ in terms of $\tau(u)$ 
\be 
r=\frac{\Lambda}{\tau^{\prime}}+\mathcal{O}\left(\Lambda^{-2}\right) .
\ee 
After some algebra using the original action we get that at the leading order at $\Lambda$ the action is given by
\be \label{eq:schT}
S_{E}=-C \int d u\left(\operatorname{Sch}(\tau, u)+\tau^{\prime 2} / 2\right), \quad C=\frac{\Phi_{r}}{8 \pi G},
\ee 
where $\operatorname{Sch}(\tau, u)$ is the so-called Schwarzian derivative given by
\be 
\operatorname{Sch}(\tau, u)=\left(\frac{\tau^{\prime \prime}}{\tau^{\prime}}\right)^{\prime}-\frac{1}{2}\left(\frac{\tau^{\prime \prime}}{\tau^{\prime}}\right)^{2} .
\ee 
After some considerations (see for example the discussion of derivation in \cite{Bagrets:2016cdf,Mertens:2017mtv,Mertens:2018fds}) one can show that the wavefunction of this quantum system\footnote{See \cite{Belokurov:2017eit,Belokurov:2019els} for path-integral considerations of Schwarzian quantum mechanics. } \eqref{eq:schT} is given by the Schrodinger equation with the Morse potential 
\be 
\left(-\partial_{\phi}^{2}+\lambda e^{\phi}+e^{2 \phi}\right) \psi_{\lambda, E}(\phi)=E^{2} \psi_{\lambda, E}(\phi).
\ee 
A similar consideration upon the change $r\rightarrow z$ for the metric given by Poincare half-plane
\be 
ds^2=\frac{1}{z^2}\left(d\tau^2+dz^2\right)
\ee 
 leads us to Liouville potential \eqref{eq:liouvillepot}.

$\,$
 
Another  class of models related to this type of potentials includes
\begin{itemize}
    \item The semiclassical limit of two-dimensional  Liouville quantum field theory leads to Liouville quantum mechanics.
    \item By analogy with the previous case, Morse quantum mechanics is the semiclassical limit of boundary Liouville quantum field theory (i.e. boundary modes of Liouville quantum gravity of constant negative curvature surfaces). Being defined on a strip  with coordinates $-\infty<\tau<\infty,\, 0 \leq \sigma \leq \pi$   boundary Liouville action has the form
    \be 
    S=\int_{-\infty}^{\infty} d \tau \int_{0}^{\pi} d \sigma\left(\frac{1}{4 \pi}\left(\partial_{a} \phi\right)^{2}+\mu e^{2 b \phi}\right)+\left.\int_{-\infty}^{\infty} d \tau M_{1} e^{b \phi}\right|_{\sigma=0}+\left.\int_{-\infty}^{\infty} d \tau M_{2} e^{b \phi}\right|_{\sigma=\pi}
    \ee 
    and in the minisuperspace limit \cite{Ghoshal:1993tm,Dorn:2006ys,Dorn:2008sw} the equations defining the wavefunction for such theory reduces to 
    \be 
\left(-\partial_{\phi}^{2}+(M_1+M_2) e^{b\phi}+\pi\mu e^{2b \phi}\right) \psi_{\lambda, k}(\phi)=E^{2} \psi_{\lambda, k}(\phi).
\ee 
\end{itemize}

To summarize,  the main object under consideration is a quantum particle a on the line in the potential of the exponential type 
\be 
V_{\lambda}(x)=\lambda\cdot \exp(x)+\exp(2x) \label{Morse_potential}
\ee 
which we call the Morse potential for simplicity and its limiting at $\lambda \to 0$ form 
\be \label{eq:liouvillepot}
V(x)=\exp(2x),
\ee 
which we call the Liouville potential. 

\subsection{Wigner function and its WKB approximation}

We consider one-dimensional Schrodinger equation
\begin{equation} \label{eq:schr}
\frac{\partial^{2} \psi}{\partial x^{2}}+\alpha^{2}[E-V (x)] \psi=0, \qquad \alpha^{2}=2 m
\end{equation}
Our main focus in  Sect.\ref{sec:W-WKB-Morse} is the calculation of the so-called Wigner function corresponding to the solution of \eqref{eq:schr} with some fixed $\lambda$ and $m$. 
Once the wavefunction of \eqref{eq:schr} is known explicitly in terms of shifted variables 
\be 
    X=\frac{1}{2}\left(x_{1}+x_{2}\right), \quad x=x_{1}-x_{2},
\ee 
one  define the Wigner function as
\be
f(X, k)=\int_{-\infty}^{+\infty} \frac{d x}{(2 \pi)} e^{-i k x } \psi\left(X+\frac{x}{2}\right) \psi^{*}\left(X-\frac{x}{2}\right).
\ee
 Given the potential $V(x)$ and the energy of the particle $E$ the WKB approximation of Wigner function can be written as
\be \label{eq:wkbgen}
f_{\mathrm{WKB}}^{0}(X, k)=\frac{2 \cos [A(X, k) / \hbar-\pi / 4]}{\pi \sqrt{\hbar}\left[\left(X+x_{0} / 2\right) p\left(X-x_{0} / 2\right) 
-\left(X-x_{0} / 2\right) p\left(X+x_{0} / 2\right)\right]^{1 / 2}},
\ee
where $A(x,k)$ with $p$ and $x_0 $ are defined below in the text. It is known for this approximation to fail to describe the wavefunction near the turning points. As it was shown in \cite{berry},  one can derive an improved version of the formula \eqref{eq:wkbgen} based on the uniform WKB approximation, which has the form
\be
\label{WKB-Berry}
f_{\mathrm{WKB}}(X,k)=\frac{\sqrt{2}[3 A(X, k) / 2]^{1 / 6} \mathrm{Ai}\left(-[3 A(X, k) / 2 \hbar]^{2 / 3}\right)}{\pi \hbar^{2 / 3}\left[\left(X+x_{0} / 2\right) p\left(X-x_{0} / 2\right) 
-\left(X-x_{0} / 2\right) p\left(X+x_{0} / 2\right)\right]^{1 / 2}}.
    \ee
It is known that the expression given by \eqref{WKB-Berry} is correctly normalized and square-integrable for bound states. An important feature of the Wigner function is that it can be defined for unbounded potentials which lead to the continuous spectrum.

The approximation  \eqref{WKB-Berry}  depends on   $A(X,k)$ and $p(X\pm x_0/2)$. The definition of these quantities  starts with the solution of  the stationary phase condition with respect to $x_0$
\bea
p(x)=\sqrt{2m\cdot(E-V(x))}\label{px}\\
p(X+x_0 / 2)+p(X-x_0 / 2)=2\cdot k,\label{x0}
\eea
and then, the term in the denominator of the Wigner function
\eqref{WKB-Berry}  is given by points in the phase space $P_1=X+x_0/2$ and $P_2=X-x_0/2$

Also we have\footnote{See \cite{Habib:1990,Habib2} for a nice exposition} $A(X, k)$, which comes from a particular integral  in the phase space and it is convenient \cite{Habib:1990,Habib2} to split it into three terms
\be 
A(X, k)=\left|A_{2}\right|+\theta\left(k_{\max }-k\right)\left|A_{3}\right|-\left|A_{1}\right|,\,\,\,\,k_{\max }=\frac{1}{2} p\left(2 X-X_{c}\right), 
\ee 
where $A_1$, $A_2$ and $A_3$ are given by
\begin{gather}
A_{1}=k x_{0}, \quad
A_{2}=\int_{X-x_{0} / 2}^{X+x_{0} / 2} d x^{\prime} p\left(x^{\prime}\right), \quad
A_{3}=2 \int_{X_{c}}^{X-x_{0} / 2} d x^{\prime} p\left(x^{\prime}\right),
\end{gather}
and $X_{c}$ is the classical turning point defined as
\begin{equation}
    E=V(X_c).
\end{equation}
As a warmup let us consider one of the simplest textbook examples of the  potential, the linear potential $V(x)=\alpha x$. This potential is unbounded and lead to continuous spectrum as well as the exponential potentials which we will consider in the next section. After some algebra all necessary information for the Wigner function can be represented in the form
\begin{gather}
  X_c=-\frac{E}{\alpha},\,\,\, x_0=\frac{2 \sqrt{2 k^2 m (\alpha  X+E)-k^4}}{\alpha  m},\,\,\,
   k_{\text{max}}=\sqrt{m(E+\alpha X)},\\
   A_1= \frac{2 k\sqrt{2 k^2 m (\alpha 
   X+E)-k^4}}{\alpha  m},\,\,\,\, A_3=\frac{4 \sqrt{2 m}}{3 \alpha}\left(E+\alpha\left(X-x_{0} / 2\right)\right)^{3 / 2},\nn\\
   A_2=\frac{2 \sqrt{2 m}}{3 \alpha}\left( \left(E+\alpha\left(X+x_{0} / 2\right)\right)^{3 / 2}-\left(E+\alpha\left(X-x_{0} / 2\right)\right)^{3 / 2}\right).
\end{gather}


\subsection{Fisher information}
Another measure of the non-classicality of the quantum system is the so-called Fisher information. If the wavefunctions of the system in the coordinate $\psi(x)$ and momentum $\tilde{\psi}(k)$ representations are known, then the Fisher information can be expressed in terms of the probability density in the coordinate $\rho(x)=\psi^*(x)\psi(x)$ and momentum $\gamma(k)=\tilde{\psi}^*(k)\tilde{\psi}(k)$ representations, respectively
\begin{gather}
    \mathcal{I}_{\rho}=\int \rho\, \partial_{x} \ln \rho(\partial_{x} \ln \rho)^* d x, \label{Fisher_rho} \\
    \mathcal{I}_{\gamma}=\int \gamma \, \partial_{k} \ln \gamma(\partial_{k} \ln \gamma)^*  d k. \label{Fisher_gamma}
\end{gather}
Also in \cite{hall} the joint measure based on both of the representations of Fisher information and called joint non-classicality has  also been introduced 
\begin{equation}
J_{n c}=\frac{\hbar}{2} \sqrt{\mathcal{I}_{\rho} \mathcal{I}_{\gamma}}\label{non-class}
\end{equation}

The simplest system where one can calculate Fisher information  is the harmonic oscillator with the potential $V(x)=1/2\omega^2 x^2$. 
The wavefunctions of harmonic oscillator in position and momentum spaces are well known
\begin{gather}
    \psi_{n}(x)=\left(\frac{m w}{\pi \hbar}\right)^{1 / 4} \frac{1}{\sqrt{2^{n} n !}} e^{-\frac{m w x^{2}}{2 \hbar}} H_{n}\left(\sqrt{\frac{m w}{\hbar}} x\right),\\
    \tilde{\psi}_{n}(k)=\frac{1}{(\pi \hbar m w)^{1 / 4}} \frac{1}{\sqrt{2^{n} n !}} e^{-\frac{k^{2}}{2 \hbar m w}} H_{n}\left(\frac{1}{\sqrt{\hbar m w}} k\right),
\end{gather}
where $H_n(x)$ is Hermite polynomial.
Probability density function for a zero-energy state:
\begin{gather}
    \rho_{0}(x)=\sqrt{\frac{m w}{\pi \hbar}} e^{-\frac{m w x^{2}}{\hbar}},  \gamma_{0}(k)=\frac{1}{\sqrt{\pi m \hbar w}} e^{-\frac{k^{2}}{m \hbar w}}.
\end{gather}
Then using definition of Fisher information (\ref{Fisher_rho}),  (\ref{Fisher_gamma}) we get
\begin{gather}
    \mathcal{I}_{\rho_0} =\frac{2mw}{\hbar},\quad  \mathcal{I}_{\gamma_0}=\frac{2}{\hbar mw},
\end{gather}
and the non-classicality is just equals to unity
\begin{equation}
    J_{nc}=1.
\end{equation}
This is the characteristic property of the nonclassicality, which is bounded by unity $J_{nc} \geq 1$ and equals unity on the simplest states like the coherent or Gaussian ones. 
It is simple to see, that for the excited states non-classicality is growing
\begin{gather}
    \mathcal{I}_{\rho_1} =\frac{6mw}{\hbar},\quad  \mathcal{I}_{\gamma_1}=\frac{6}{\hbar mw},\quad
    J_{nc_1}=3,\\
    \mathcal{I}_{\rho_2} =\frac{10mw}{\hbar},\quad  \mathcal{I}_{\gamma_2}=\frac{10}{\hbar mw},\quad
    J_{nc_2}=5,
\end{gather}
where subscripts ``1'' and ``2'' correspond to the first and second excited states.
Calculating information for unbounded potentials, which are characterized by a continuous spectrum, faces the difficulty of taking integrals at infinity because the wavefunctions oscillate like a plane wave. To solve this problem, we introduce regularization. Let's put an infinite vertical wall at the point $x=-L$. Then we get potential with a well and a discrete spectrum, and the wavefunctions will exponentially decrease by $x\to \pm \infty$. 
Let's consider the above in a concrete example.

\section{Wigner function for  black hole}\label{sec:W-WKB-Morse}
In the previous section we described in general terms some classes of two-dimensional quantum gravity models which are effectively reduce to the particular type of potentials with one or exponents. To establish the notation we consider one-dimensional Schrodinger equation
\begin{equation} \label{eq:schr}
\frac{\partial^{2} \psi}{\partial x^{2}}+\alpha^{2}[E^2-V_\lambda (x)] \psi=0,
\qquad
\alpha^{2}=2 m
\end{equation}
and $V_\lambda(x)$ is (\ref{Morse_potential})

The wavefunction  for a general $\lambda$ and $\alpha=1$ in position has the form
\begin{gather}
\psi(x)={\cal N}_E \cdot  e^{-e^{x }+i E x} \,U\left(i E+\frac{\lambda }{2}+\frac{1}{2},2 i E+1,2 e^{x }\right) \label{eq:confl}
\end{gather} 
where $U(a, b, z)$  is the confluent Hypergeometric function\footnote{ The confluent Hypergeometric function  can be represented as an integral
$
U(c, b, z)=\frac{1}{\Gamma(c)} \int_{0}^{\infty} e^{-z t} t^{c-1}(1+t)^{b-c-1} d t,$ where $ \operatorname{Re} c>0
$
with $\lambda>-1$ ($k$ belongs reals).}  and ${\cal N}_E$ is the normalization to be chosen, $E>0$. wavefunction in momentum space could be find by taking Fourier transform of (\ref{eq:confl}):
\begin{gather}
    \tilde{\psi}(k)=\frac{1}{\sqrt{2\pi}}\int^{+\infty}_{-\infty} \psi(x)e^{-ikx}dx=\frac{2^{-\frac{1}{2}-i E} \Gamma (i (E-k))
   }{\sqrt{\pi } \Gamma
   \left(\frac{\lambda }{2}+i E+\frac{1}{2}\right)}\mathcal{N}_{E}\cdot \nn\\
     \left(\frac{2^{-\frac{\lambda }{2}-\frac{1}{2}} \Gamma
   \left(-i k-\frac{\lambda }{2}-\frac{1}{2}\right) \Gamma
   \left(\frac{\lambda}{2} + i E+\frac{1}{2}\right) \,
   _2F_1\left(\frac{\lambda}{2} - i E+\frac{1}{2}),\frac{\lambda}{2}
   + i E+\frac{1}{2}; i k+\frac{\lambda}{2}
   +\frac{3}{2});\frac{1}{2}\right)}{\Gamma (i (E-k))}\right.+\nn\\
   \left.\frac{2^{i k}
   \Gamma (-i(k+E)) \Gamma \left( i k+\frac{\lambda}{2}
   +\frac{1}{2})\right) \, _2F_1\left(i (E-k),-i (k+E);- i
   k-\frac{\lambda}{2} +\frac{1}{2});\frac{1}{2}\right)}{\Gamma \left(
   \frac{\lambda}{2} - i E+\frac{1}{2})\right)}\right)
\end{gather}

\begin{figure}[h!]
\centering 
\includegraphics[width=8.cm]{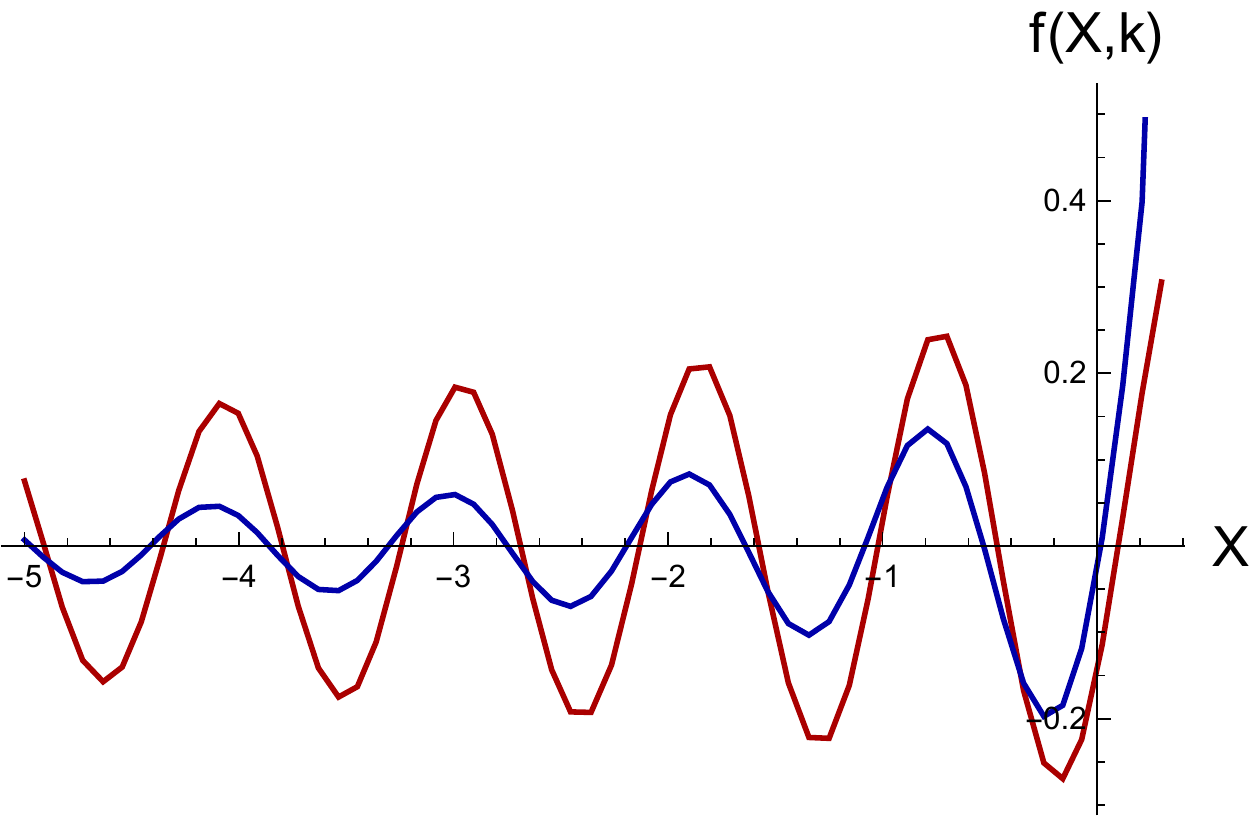}
 \caption{Wigner function  defined by \eqref{eq:wkbgen} and \eqref{eq:morsesol} for fixed $k=2$ (blue curve) and the red one to $k=1.5$ (red curve). The energy are fixed to be $E=4$ and $\lambda=0.1$ for both plots.}
 \label{fig:fX}
\end{figure}

The function \eqref{eq:confl} has the explicit but  complicated form and this  does not allow us to find the Wigner function explicitly. To circumvent this issue we will use WKB approximation which is proven to be useful in many physical applications. 
For the Morse potential stationary phase condition (\ref{x0}) has the form
\bea \label{eq:x0morse}
   \sqrt{m \left(E-\lambda e^{X-\frac{x_0}{2}}
   -e^{2
   \left(X-\frac{x_0}{2}\right)}\right)}+
   \sqrt{m \left(E-\lambda e^{X+\frac{x_0}{2}} -e^{2
   \left(X+\frac{x_0}{2}\right)}\right)}=
    \sqrt{2} k\nn\\\label{x0M}\eea
   The  turning point corresponding to Morse potential has the form
   \be \label{CTP-Morse}
   X_c=\log \left(\frac{1}{2} \left(\sqrt{4
   E+\lambda^2}-\lambda\right)\right)\ee
   with the expressions for $A_i, i=1,2,3$ given explicitly by
   \begin{gather} \label{eq:morsesol}
   A_1=kx_0 \\
   A_2=p\left(X+\frac{x_0}{2}\right)-p\left(X-\frac{x_0}{2}\right)-\nn\\
   \frac{\lambda  \sqrt{m} }{\sqrt{2}}\text{arctan}\left(\frac{\sqrt{m}(\lambda +2 e^{X+\frac{x_0}{2}})}{\sqrt{2}p\left(X+\frac{x_0}{2}\right)}\right)+\frac{\lambda 
   \sqrt{m} }{\sqrt{2}}\text{arctan}\left(\frac{\sqrt{m}(\lambda +2 e^{X-\frac{x_0}{2}})}{\sqrt{2}p\left(X-\frac{x_0}{2}\right)}\right)+\nn\\
   \sqrt{2Em} \, \text{artanh}\left(\frac{\sqrt{m}(\lambda  e^{X+\frac{x_0}{2}}-2 E)}{
   \sqrt{2E}  p\left(X+\frac{x_0}{2}\right)}\right)-
    \sqrt{2Em}\,
   \text{artanh}\left(\frac{\sqrt{m}(\lambda  e^{X-\frac{x_0}{2}}-2 E)}{ \sqrt{2E}
   p\left(X-\frac{x_0}{2}\right)}\right)\\
   A_3=-\frac{\pi  \sqrt{m} \left(\lambda +2 i \sqrt{E}\right)}{
   \sqrt{2}}+2p\left(X-\frac{x_0}{2}\right)-\nn\\
   \lambda 
   \sqrt{2m}\,\text{arctan}\left(\frac{\sqrt{m}(\lambda +2 e^{X-\frac{x_0}{2}})}{\sqrt{2}p\left(X-\frac{x_0}{2}\right)}\right)+2\sqrt{2Em}\,
   \text{artanh}\left(\frac{\sqrt{m}(\lambda  e^{X-\frac{x_0}{2}}-2 E)}{ \sqrt{2E}
   p\left(X-\frac{x_0}{2}\right)}\right)\\
   k_{\text{max}}=\sqrt{-\frac{m \left(E+e^{2 X}\right) \left(-2 E+\lambda 
   \left(\sqrt{4 E+\lambda ^2}-\lambda \right)+2 e^{2
   X}\right)}{\left(\lambda -\sqrt{4 E+\lambda ^2}\right)^2}}
\end{gather}
The solutions of the equation \eqref{x0M} defining $x_0$ can be found numerically and we present the plot for Wigner function corresponding to the two-dimensional quantum boundary Liouville/Schwarzian quantum mechanics in Fig.\ref{fig:fX}.

\section{Fisher information for regularized JT black hole}
\label{sec:fisher}
\begin{figure}[h]
\begin{minipage}[h]{0.45\linewidth}
\center{\includegraphics[width=1\linewidth]{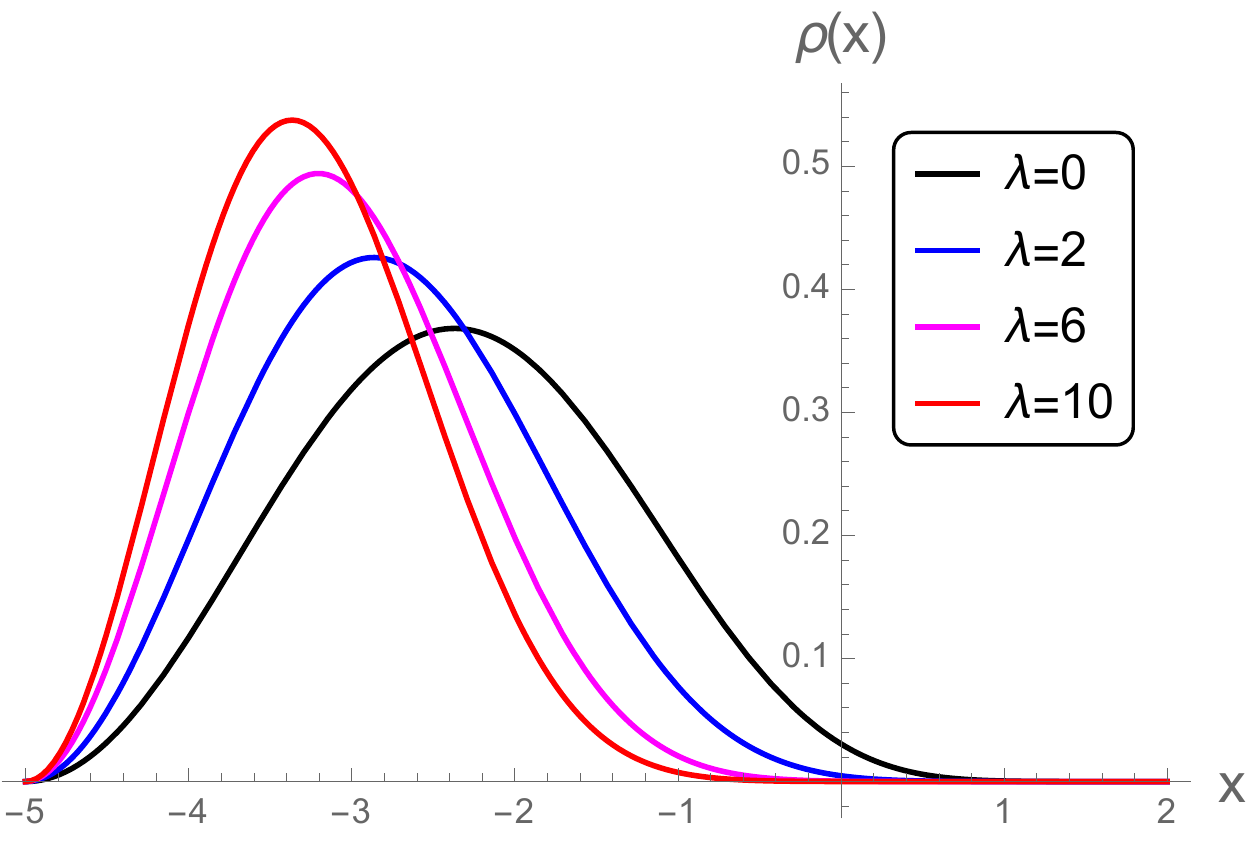} \\ a)}
\end{minipage}
\hfill
\begin{minipage}[h]{0.45\linewidth}
\center{\includegraphics[width=1\linewidth]{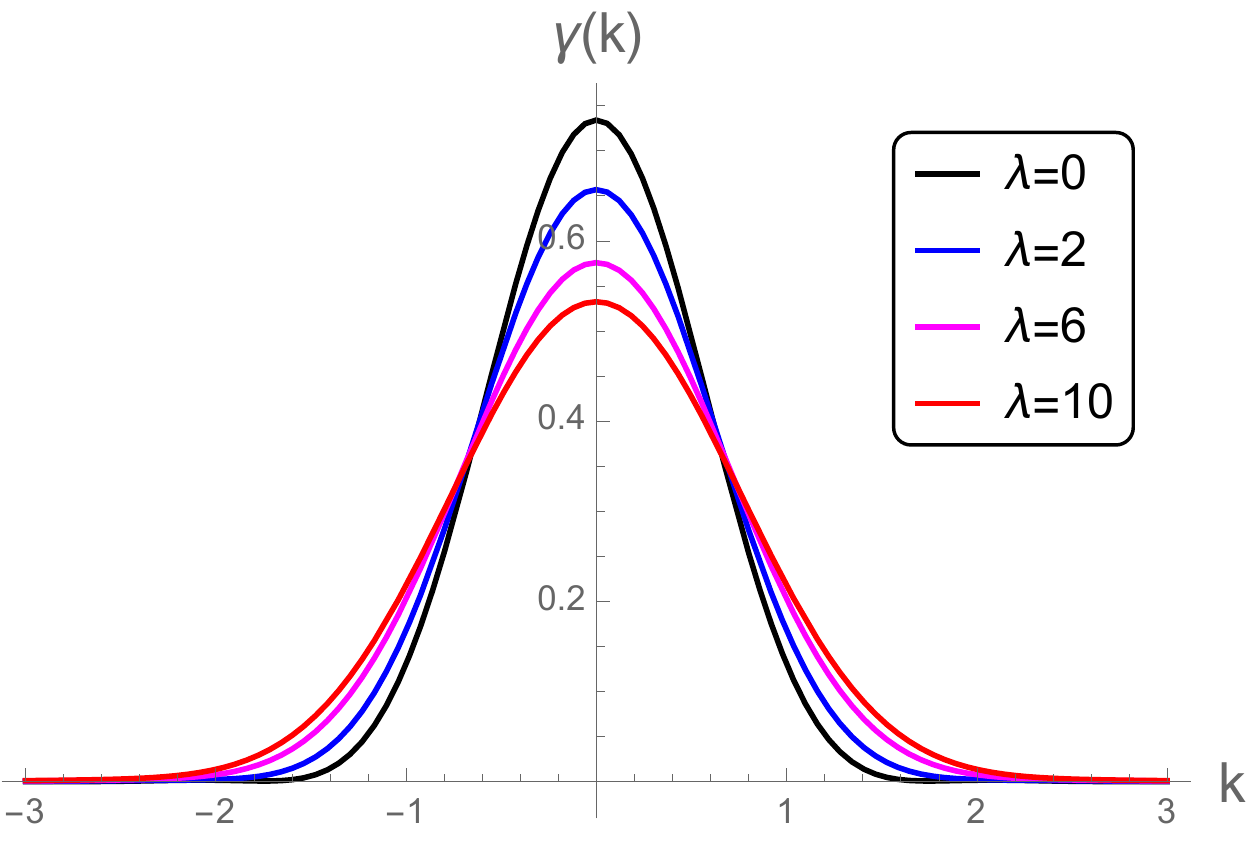} \\ b)}
\end{minipage}
\caption{The probability density  for the ground state of particle in regularized Morse potential  in  the position (left plot) and momentum representation (right plot).  Different colors correpond to $\lambda=0$ (black curve), $\lambda=2$ (blue curve), $\lambda=6$ (magenta curve) and $\lambda=10$ (red curve).  Here $L=5$.}
\label{fig:image1}
\end{figure}

\begin{figure}[h]
\begin{minipage}[h]{0.47\linewidth}
\center{\includegraphics[width=1\linewidth]{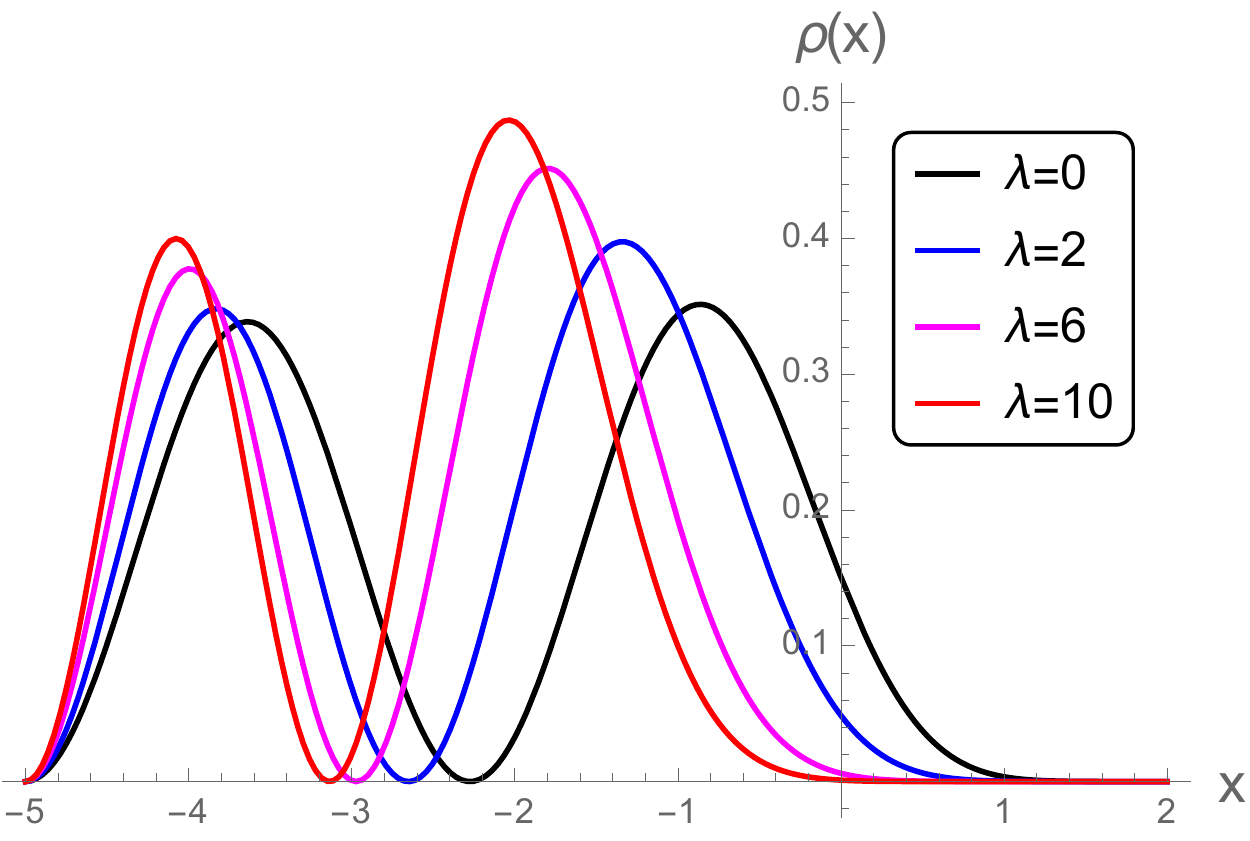} \\ a)}
\end{minipage}
\hfill
\begin{minipage}[h]{0.49\linewidth}
\center{\includegraphics[width=1\linewidth]{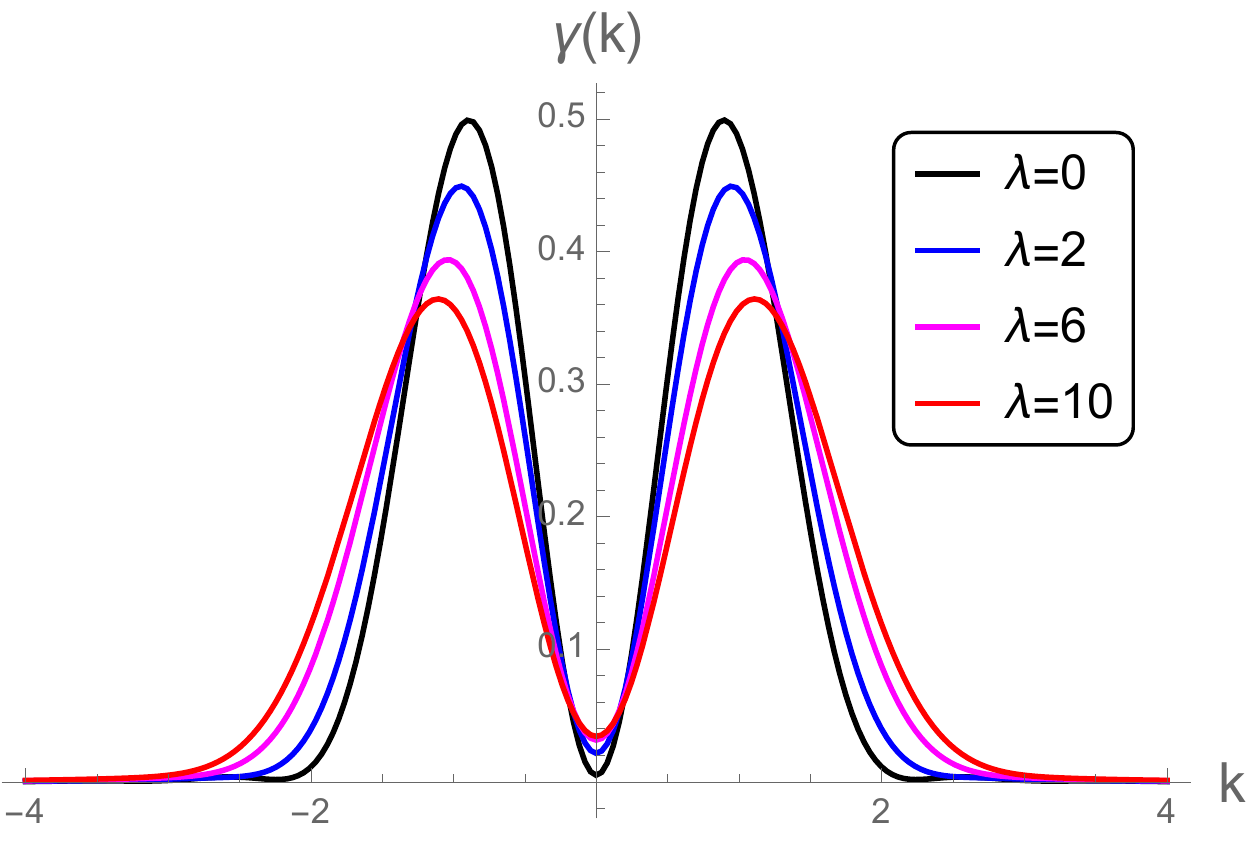} \\ b)}
\end{minipage}
\caption{The probability density  for the ground state of particle in regularized Morse potential  in  the position (left plot) and momentum representation (right plot).  Different colors correpond to $\lambda=0$ (black curve), $\lambda=2$ (blue curve), $\lambda=6$ (magenta curve) and $\lambda=10$ (red curve).  Here $L=5$.}
\label{fig:image2}
\end{figure}

As it was mentioned before, the Fisher information for unbounded potential is  ill-defined. To proceed further and calculate non-classiality and Fisher information for black hole (i.e. unbounded Morse potential) we introduce the regularization imposing the boundary condition, which terminates the wavefunction at distance $L$
\be 
\psi(-L)=0.
\ee 
This  leads  to the discrete spectrum and the quantization condition on $E$ 
\begin{gather}
    U\left(\frac{1}{2} i (2 E_n-i \lambda-i),1+2 i E_n,2 e^{-L}\right)=0. \label{boundary_conditional}
\end{gather}
Finding  the spectrum $E_n$ and the normalization constant numerically for a particular regularization $L=5$ we calculate the Fisher information in position and momentum space and the joint non-classicality for the ground and first excited states (see Fig.\ref{fig:image1} and Fig.\ref{fig:image2} for plots of the wavefunctions). 

\begin{figure}[h!]
\centering
\includegraphics[width=5cm]{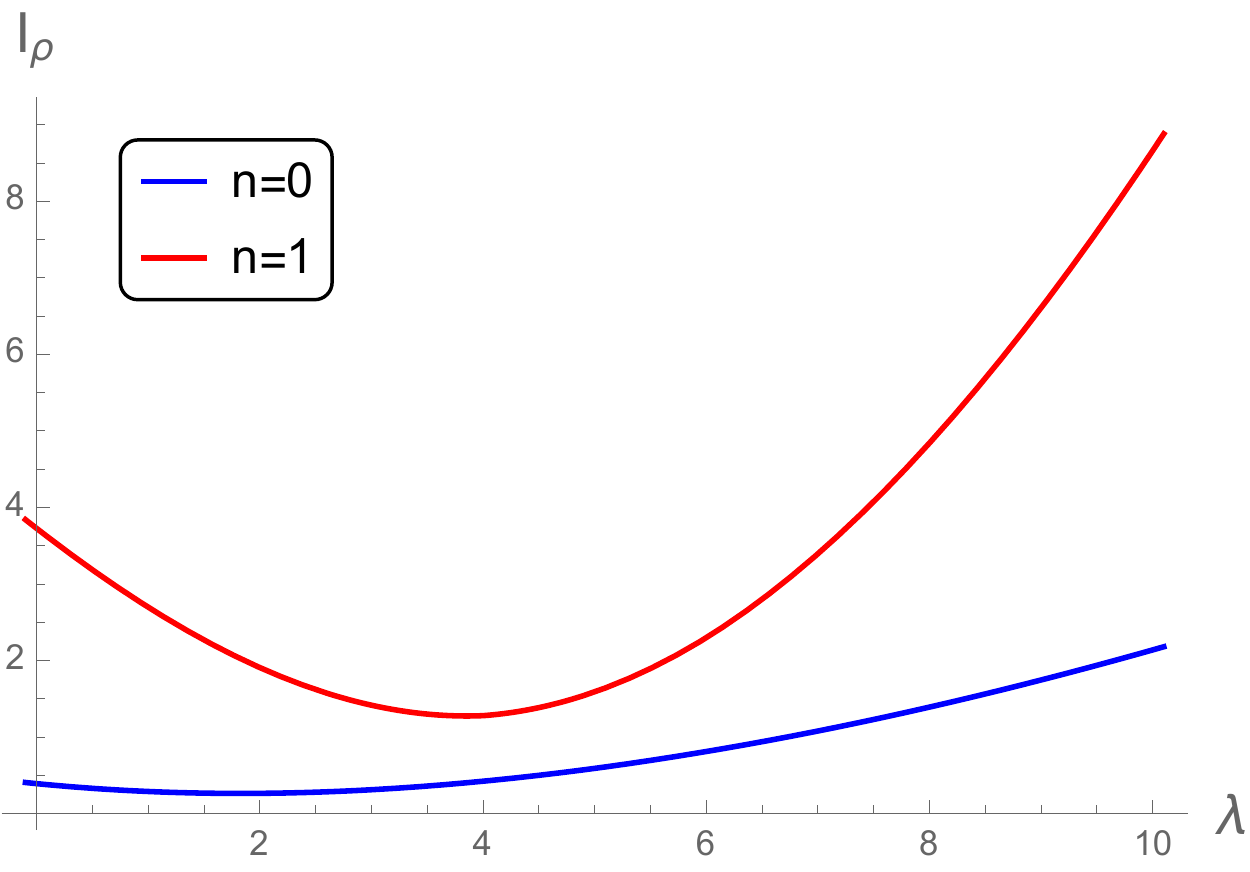}
\includegraphics[width=5cm]{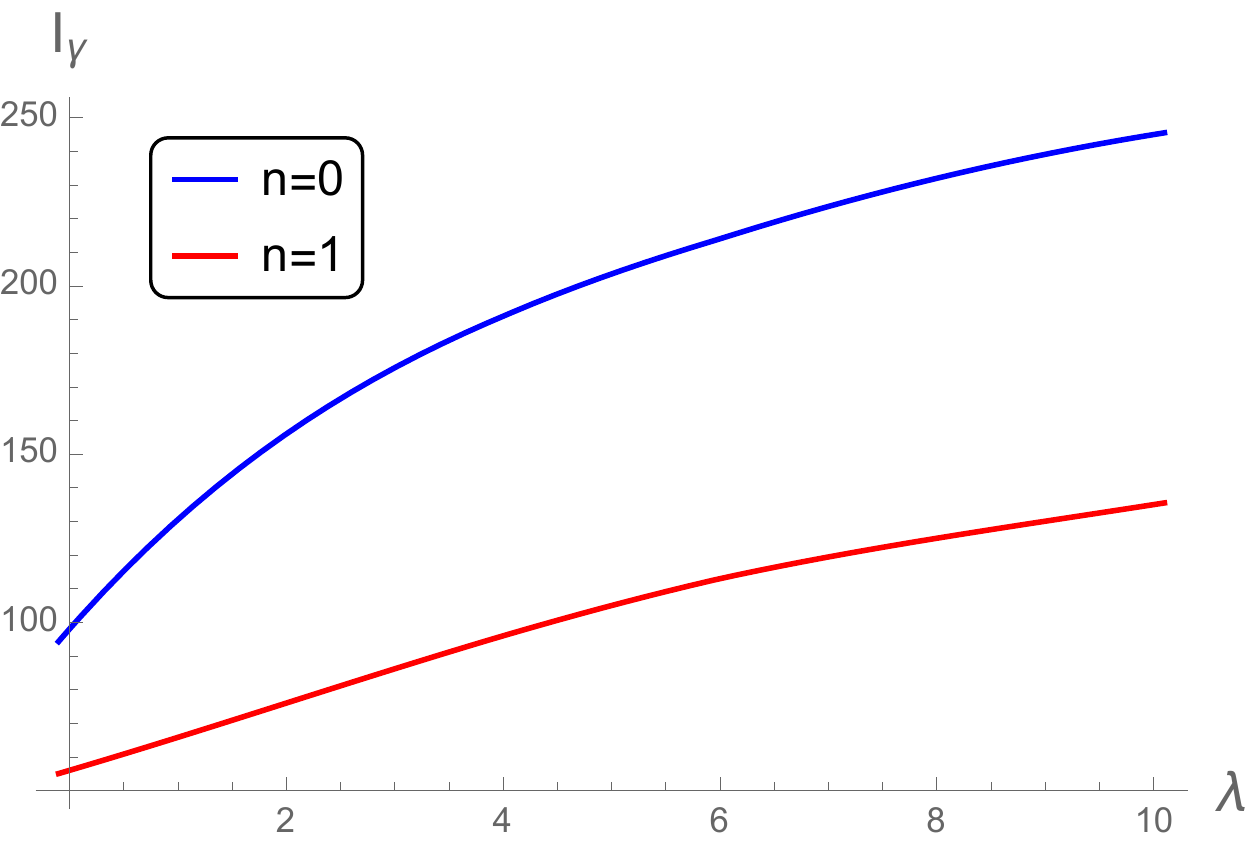}
\includegraphics[width=5cm]{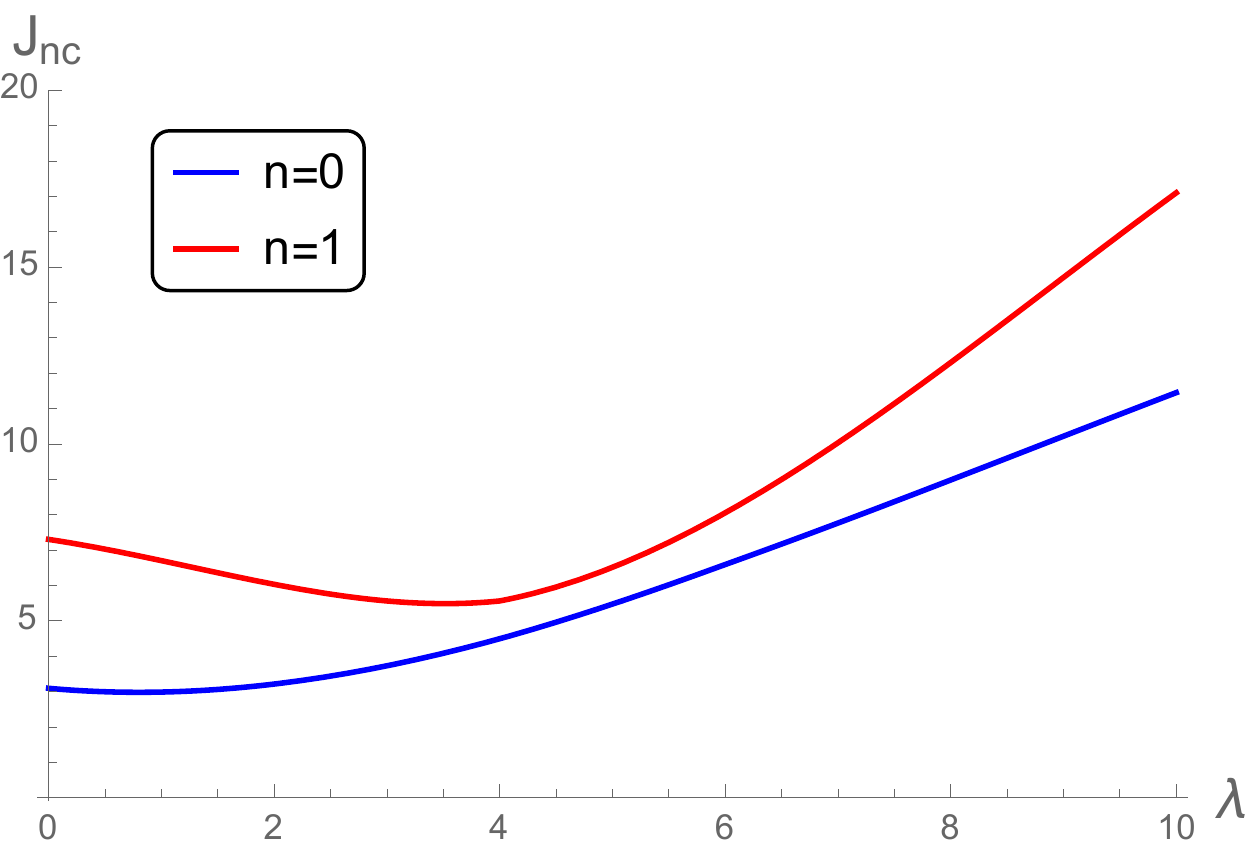}\\
a)\qquad \qquad\qquad \qquad\qquad b)\qquad
\qquad
\qquad\qquad\qquad c)
 \label{fig:fisher}
 \caption{ The dependence of the Fisher information corresponding to the regularized $\qquad $($L=5$) Morse potential in position a)  and momentum space b) on $\lambda$. c) the same dependence for joint nonclassicality.}
\end{figure}

\newpage
We  present the results of the calculation of Fisher information in Fig.\ref{fig:fisher}. Notice, that as for the harmonic oscillator  position and momentum Fisher informations are of different order of magnitude.

\newpage
\section*{Conclusion}
To summarize, we have calculated Fisher information and Wigner function for a system describing black holes in two-dimensional quantum gravity. It would be interesting to understand how the Fisher information and Wigner function approaches could be incorporated in the different holographic proposals of Fisher metric and minisuperspace approximation to $AdS$ gravity considered in the literature recently \cite{Caputa:2018asc,Banerjee:2017qti,MIyaji:2015mia,Erdmenger:2020vmo}. Another interesting direction to consider is to understand better the continuous spectrum definition for the Fisher information and the exact calculation for the non-regularized quantum mechanics, as well as for baby universes and wormhole constructions.

\section*{Acknowledgements}
This work is supported by the ``Basis''
Science Foundation (grant No. 18-1-1-80-4).

\newpage

\end{document}